\documentclass[aps, pra, twocolumn, notitlepage, 10pt, superscriptaddress
]{revtex4-2}

\usepackage{graphicx}
\usepackage{dcolumn}
\usepackage{bm}
\usepackage[percent]{overpic}
\usepackage{amsmath}
\usepackage{xcolor}

\begin{document}


\title{
Collective dynamics of active matter with orientation-weighted alignment}

\author{Bohdan Dobosh}
\affiliation{Department of Physics, Taras Shevchenko National University of Kyiv, 64/13, Volodymyrska Street, Kyiv 01601, Ukraine}

\author{Alexander Yakimenko}
\affiliation{Department of Physics, Taras Shevchenko National University of Kyiv, 64/13, Volodymyrska Street, Kyiv 01601, Ukraine}
\affiliation{Dipartimento di Fisica e Astronomia Galileo Galilei,
Universit\'a di Padova,
Via Marzolo 8, 35131 Padova, Italy}

\date{\today}

\begin{abstract}
We study an agent-based model of self-propelled particles with a velocity-dependent alignment rule. This interaction is orientation weighted and acts along the line connecting neighboring particles. Tuning the alignment strength produces several distinct collective regimes, including disordered gas-like motion, coherent flocking, jammed high-density states, and densely ordered moving clusters with active-crystal-like behavior. These results show that a simple local alignment rule can generate a broad range of nonequilibrium collective dynamics within a single microscopic model.
\end{abstract}

\maketitle

\section{Introduction}

Active matter is a class of nonequilibrium systems composed of self-propelled particles or agents that continuously convert internal or environmental energy into directed motion and thus generate collective behavior far from thermal equilibrium~\cite{annurev_condmatphys,FODOR2018106,Marchetti2013}. Examples range from bacterial suspensions, cell layers, and animal groups to synthetic systems such as colloidal rollers, magnetic colloids, vibrated granular media, and active nematics~\cite{Sanchez2012,bricard2013nature,bricard2015natcomm,kaiser2017sciadv,blair2003pre,narayan2006science,Active_turbulence,mecke2023ncomms}. Simple local interactions in such systems can produce robust large-scale collective behavior, including flocking, clustering, jamming, and dense ordered motion~\cite{Vicsek1995,TonerTu1998,Shaebani2020,Baconnier2025}.

Local alignment is one of the main mechanisms underlying collective motion in active matter. The Vicsek model and its extensions showed that short-range alignment can generate polar order and nonequilibrium transitions in systems of self-propelled particles~\cite{Vicsek1995,TonerTu1998}. When alignment is combined with attractive or repulsive interactions, the collective behavior becomes richer, leading to cohesive flocks, clustered states, and dense moving aggregates~\cite{Gregoire2003,Das2020}. Related phenomena have also been observed in synthetic active matter. Colloidal rollers display flocking and vortex formation~\cite{bricard2013nature,bricard2015natcomm}, while active colloids can form long-lived dense structures, including living crystals and other stable collective states~\cite{Palacci2013,Mognetti2013,Bauerle2020}. 
At the continuum and mesoscopic levels, dense active motion has been described in terms of active crystals, active solids, and dense active-matter models of collective motion~\cite{Ferrante2013ActiveCrystals,MenzelLoewen2013,Menzel2014,Henkes2020Dense}, and two-dimensional active systems are now known to exhibit a broad range of liquid, hexatic, solid, and phase-separated regimes~\cite{Klamser2018,cates2015annurev,Das2022,ActiveLesHouchesCorr}.

These developments motivate the search for simple microscopic interaction rules that can generate different collective regimes within a single dynamical framework. Most Vicsek-type alignment models prescribe alignment through angular averaging of neighboring headings, although many extensions modify the interaction symmetry, include cohesion, or add additional force-based ingredients~\cite{Vicsek1995,Chate2008Variations,Shaebani2020,Baconnier2025}. Here we instead introduce a force-based alignment rule: the contribution from a neighbor acts along the line connecting the two particles, while its strength and sign are determined by the projection of the neighbor's velocity onto this line. Since this projection depends on both the neighbor's heading and speed, the response is controlled not only by how the neighbor moves, but also by where it is located relative to the particle under consideration~\cite{Shaebani2020,Baconnier2025,Xiao2024MotionSalience}.

This rule differs from heading-averaging alignment, and also from many force-based cohesive or selective-response models in which attraction--repulsion, escape--pursuit, and alignment mechanisms are introduced as separate prescribed interactions~\cite{Gregoire2003,Romanczuk2009EscapePursuit,Romanczuk2012Selective,Shaebani2020,Baconnier2025}. In the present model, the effective alignment force is tied directly to the velocity projection on the interparticle axis. The same neighbor can therefore generate an acceleration either toward or away from it, and can either increase or decrease the particle speed, depending on the relative position and direction of motion. This geometry-dependent coupling introduces a front--back asymmetry, absent in the minimal Vicsek alignment rule, and directly couples spatial organization, acceleration, and alignment. The model therefore provides a compact setting for studying how anisotropic velocity-dependent interactions can produce flocking, dispersive motion, dense ordering, and collective turning within the same microscopic description.

Such anisotropic response is also relevant to biological collectives. 
Experiments on bird flocks and insect groups have shown that directional changes and collective responses to perturbations can propagate through a group via local interactions and behavioral inertia, without requiring designated leaders~\cite{Attanasi2014StarlingTurns,Escudero2010DirectionalSwitching,vanderVaart2019SwarmSpectroscopy,Zheng2024BOC}.
More recent work has further emphasized that the response to neighbors can depend on relative position and directional perception, rather than on heading alone~\cite{Xiao2024MotionSalience,Castro2024OpticFlow}. This motivates studying models in which coherent reversals arise from the interaction rule itself and from the response of the group to localized external forcing.

In this work we introduce an agent-based model of self-propelled particles with three ingredients: orientation-weighted alignment, short-range Lennard-Jones interactions, and speed saturation. In the absence of external forcing, the sign of the alignment strength separates coherent flocking from dispersive gas-like motion. We then use an anisotropic target field, constructed in analogy with the same orientation-weighted interaction, to probe how an ordered group responds to spatially localized cues. Depending on the target strength and alignment parameters, the system evolves toward either jammed high-density states or densely ordered moving clusters with local sixfold structure and coherent reversals. The model therefore provides a compact framework for studying how geometry-dependent alignment, short-range interactions, and anisotropic forcing shape nonequilibrium collective dynamics.

The paper is organized as follows. Section~II introduces the model and simulation procedure. Section~III presents the main collective regimes and discusses stochasticity and effective friction. Section~IV summarizes the results.

\section{Model}
We consider a system of \(N\) self-propelled agents moving in two spatial dimensions. Time is discretized with step \(\Delta t\), while particle positions remain continuous. The dynamics are described by
\begin{equation}
    \ddot{\mathbf{r}}_i = \mathbf{a}_i^{\mathrm{int}} + \mathbf{a}_i^{\mathrm{bord}} + \mathbf{a}_i^{\mathrm{fl}},
    \label{eq:a_tot}
\end{equation}
where \(\mathbf{r}_i\), \(\mathbf{v}_i\), and \(\mathbf{a}_i\) denote the position, velocity, and acceleration of the \(i\)-th agent. The three terms on the right-hand side represent interparticle interactions, confinement by the boundary, and alignment-driven motion, respectively.
\begin{figure}[t]
\includegraphics[width=\linewidth]{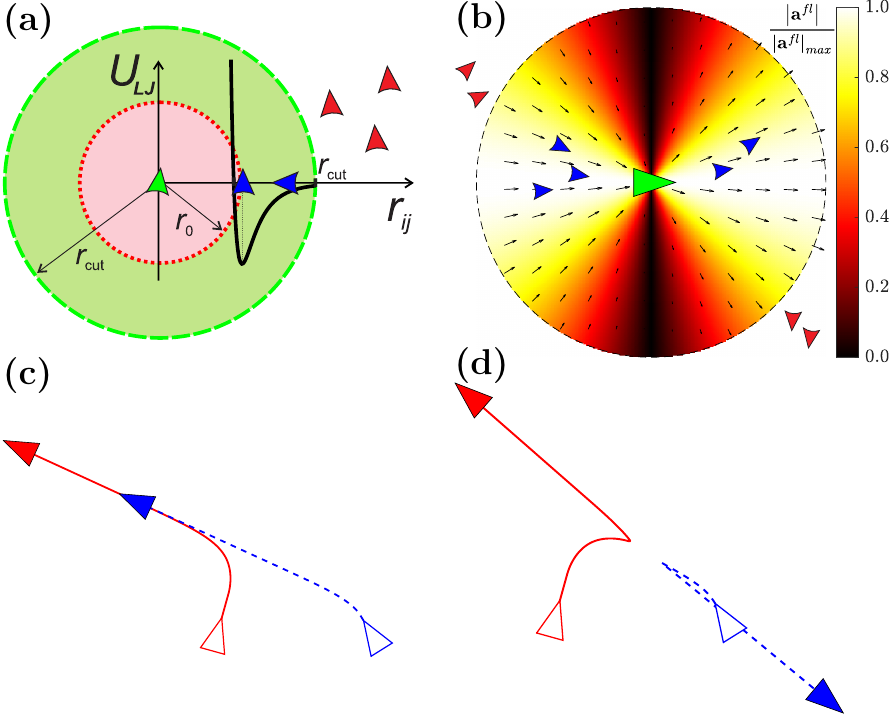}
    \caption{    
(a) Lennard-Jones-type pair potential associated with a central agent (green triangle). The red-shaded region corresponds to short-range repulsion that prevents collisions, while the green-shaded region indicates the attractive part of the interaction.
(b) Alignment force field generated by a single agent located at the center (green triangle). Arrows show the force direction, and the background color represents the absolute value of the force.
Panels (c) and (d) illustrate the dynamics of a two-agent system with the same initial conditions, shown by empty red and blue triangles, but opposite signs of the alignment parameter \(A\); the corresponding trajectories are shown by solid red and dashed blue lines.
(c) For \(A>0\), the agents become aligned and move with a common velocity.
(d) For \(A<0\), the agents first become anti-aligned and eventually repel each other.}
    \label{fig:flocking}
\end{figure}

Pair interactions are described by a truncated Lennard-Jones-type potential, shown in Fig.~\ref{fig:flocking}(a). The parameter \(r_{\rm cut}\) defines the interaction range: only particles with \(r_{ij}=|\mathbf{r}_j-\mathbf{r}_i|\le r_{\rm cut}\) contribute to the pair potential and alignment force. The interaction is repulsive at short distances, attractive at intermediate distances, and has a minimum at the equilibrium separation \(r_0\). For \(r_{ij}\le r_{\mathrm{cut}}\), we write
\begin{equation}
\label{eq:U_LJ}
U_{LJ}(r_{ij})=
U_p\left[
\left(\frac{r_0}{r_{ij}}\right)^{12}
-2\left(\frac{r_0}{r_{ij}}\right)^6
\right]
+U_{\mathrm{shift}},
\end{equation}
while \(U_{LJ}(r_{ij})=0\) for \(r_{ij}>r_{\mathrm{cut}}\). Here \(U_p\) sets the interaction strength, and the constant shift \(U_{\mathrm{shift}}\) is chosen so that \(U_{LJ}(r_{\mathrm{cut}})=0\).

The corresponding interaction acceleration acting on the \(i\)-th agent is
\begin{equation}
\label{eq:a_int_2}
\mathbf{a}_i^{\mathrm{int}}
=
-\nabla_i \sum_{j\ne i} U_{LJ}(r_{ij}),
\end{equation}
where \(\nabla_i \equiv \partial/\partial \mathbf{r}_i\) denotes the gradient with respect to the coordinates of the \(i\)-th agent.

To confine the system to a circular arena of radius \(R_b\), we introduce a boundary force
\(\mathbf{a}_i^{\mathrm{bord}}=-\nabla_i U_i^{\mathrm{bord}}\). The corresponding potential is flat inside the arena and becomes parabolic only outside it, so that agents move freely for \(r_i\le R_b\) and are pushed back only when they cross the boundary:
\begin{equation}
\label{eq:U_bord}
    U_i^{\mathrm{bord}}=\frac{k}{2}(r_i-R_b)^2\,\Theta(r_i-R_b),
\end{equation}
where \(r_i=|\mathbf{r}_i|\), \(k\) is the boundary stiffness, and \(\Theta(x)\) is the Heaviside step function.

In addition to the Lennard-Jones force \(\mathbf{a}_i^{\mathrm{int}}\) and the boundary force \(\mathbf{a}_i^{\mathrm{bord}}\), each agent is subject to an orientation-weighted alignment interaction generated by its neighbors. For a given neighbor \(j\), the corresponding contribution acts along the line connecting the two particles, i.e. parallel to the unit vector \(\hat{\mathbf r}_{ij} = (\mathbf{r}_j - \mathbf{r}_i)/r_{ij}  \), and its sign and strength are determined by the projection \((\hat{\mathbf r}_{ij}\cdot \mathbf v_j)\) of the neighbor's velocity onto this direction. We therefore write
\begin{equation}
    \label{eq:a_fl}
    \mathbf{a}_i^{\mathrm{fl}}
    =
    A\sum_{j\neq i}\hat{\mathbf r}_{ij}
    (\hat{\mathbf r}_{ij}\cdot \mathbf v_j)\Theta(r_{\mathrm{cut}}-r_{ij}),
\end{equation}
where \(A\) sets the interaction strength. Positive \(A\) favors alignment, while negative \(A\) leads to counter-alignment. Consequently, \(|A|^{-1}\) defines the characteristic alignment time scale.

The action of this term for \(A>0\) is illustrated in Fig.~\ref{fig:flocking}(b). To interpret the force field, we consider the contribution generated by a single central agent, shown in green. The interaction is strongest along the agent’s direction of motion: particles located behind the green agent are pulled toward it, whereas particles in front are pushed away. Its magnitude decreases with increasing angular deviation from this direction and vanishes in the perpendicular direction. In the many-body system, these pairwise contributions simply add. As a result, already aligned particles tend to accelerate in the same direction, while the combined action of the alignment term and the short-range Lennard-Jones interaction favors the formation of co-moving pairs and flocks.

For \(A<0\), the alignment term favors counter-alignment, as illustrated in Fig.~\ref{fig:flocking}(d). In a typical two-agent encounter, the particles acquire nearly antiparallel velocities along the line connecting them, approach head-on, collide, and are then driven apart by the short-range Lennard-Jones repulsion. In the many-body system, repeated encounters of this type suppress coherent flocking and lead to the gas-like regime discussed in Sec.~III.


We integrate the equations of motion using the velocity Verlet algorithm. For all agents, the update is performed simultaneously according to
\begin{align}
    \label{eq:Verlet1}
    \mathbf{v}(t+\frac{1}{2}\Delta t) &= \mathbf{v}(t) + \frac{1}{2}\mathbf{a}(t)\,\Delta t, \\
    \label{eq:Verlet2}
    \mathbf{r}(t+\Delta t) &= \mathbf{r}(t) + \mathbf{v}(t+\frac{1}{2}\Delta t)\,\Delta t, \\
    \label{eq:Verlet3}
    \mathbf{v}(t+\Delta t) &= \mathbf{v}(t+\frac{1}{2}\Delta t) + \frac{1}{2}\mathbf{a}(t+\Delta t)\,\Delta t .
\end{align}
In the conservative limit, i.e. for $A=0$, this reduces to the standard second-order symplectic velocity Verlet scheme. It is therefore a natural choice, providing good stability and accurate integration of the interaction dynamics while allowing the active contribution to be incorporated straightforwardly.

\begin{figure}
\includegraphics[width=\linewidth]{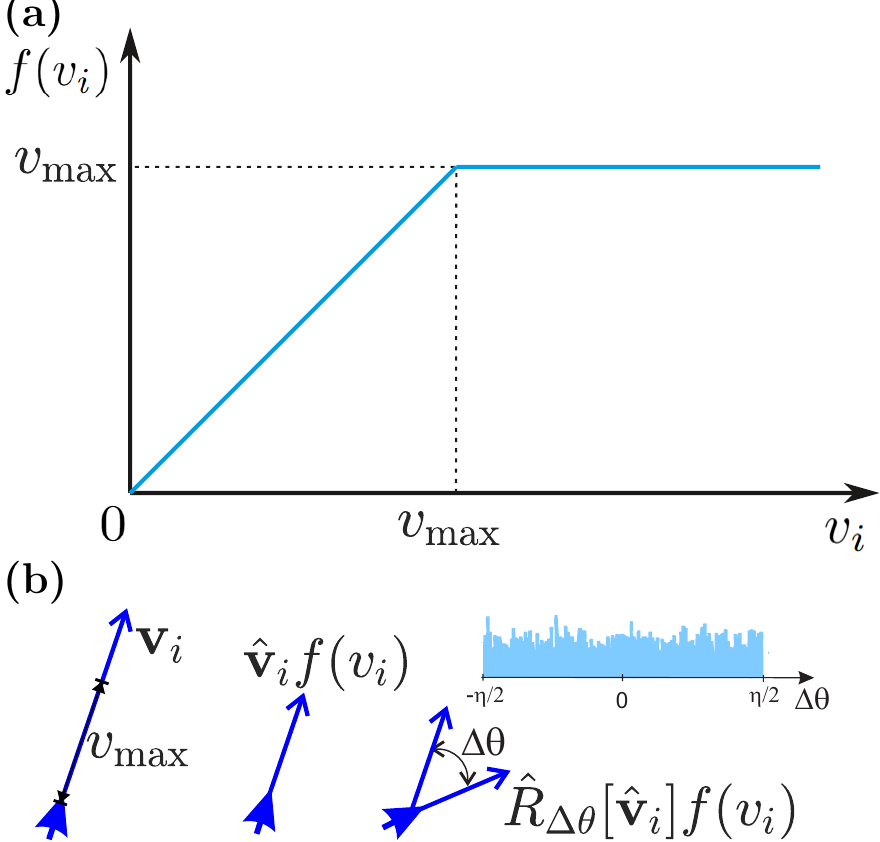}
    \caption{ (a) Speed truncation function $f: v_i \mapsto f(v_i)$. (b) Schematics of velocity rescaling, discussed in Section III: speed truncation and rotation by angle $\Delta \theta$. $\hat{\mathbf{v}}_i = \mathbf{v}_i / |\mathbf{v}_i|$.}
    \label{fig:vmax}
\end{figure}

Because the alignment term can continuously accelerate particles, we impose a speed-saturation rule and introduce a maximum speed \(v_{\max}\). This rule mimics finite propulsion capacity and prevents unbounded growth of particle speeds, while retaining nontrivial speed dynamics within the flock~\cite{Cavagna2022MarginalSpeed}.
 After each velocity update step [Eqs.~(\ref{eq:Verlet1}) and (\ref{eq:Verlet3})], the speed is truncated as illustrated in Fig.~\ref{fig:vmax}(a): if \(|\mathbf v_i|>v_{\max}\), it is reset to \(v_{\max}\), whereas lower speeds are left unchanged. Aligned particles therefore tend to accelerate until they approach this limiting speed, as illustrated in Fig.~\ref{fig:flocking}(c). Simulations without Lennard-Jones interactions confirm that this acceleration is produced by the alignment term itself.

Note that the speed cap should be distinguished from a fixed-speed update. In the simple collinear two-particle limit, an agent with initial speed \(v_0<v_{\max}\) accelerates under the alignment term approximately as \(v(t)\simeq v_0\exp(At)\) until the cap is reached. The corresponding saturation time is \(t_{\rm sat}\simeq A^{-1}\ln(v_{\max}/v_0)\). For the representative parameters used below, \(\Delta t=0.01\ll |A|^{-1}\), so the acceleration toward \(v_{\max}\) is resolved in time. The agents therefore retain variable-speed dynamics on a short timescale, even though long-lived aligned motion typically occurs close to the limiting speed.

\begin{figure}[t]
    \centering
\includegraphics[width=\linewidth]{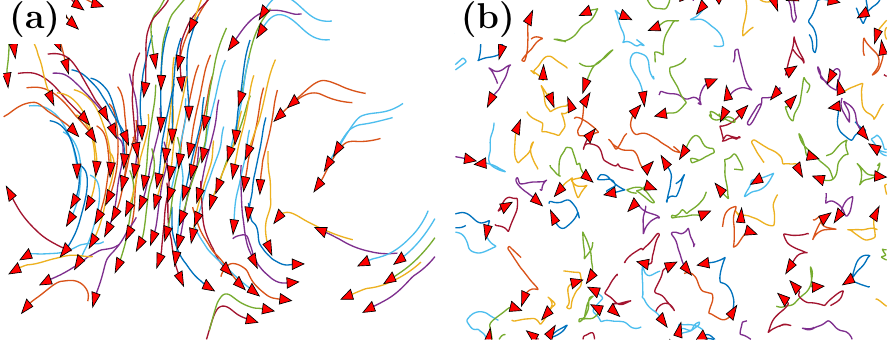}
    \caption{
    The fragments of the evolution of the system in two different regimes in the absence of external potentials. Agents are depicted with red triangles, the colored lines represent a fragment of their trajectories.
(a) Flocking regime $A>0$.
(b) Active gas $A<0$. Detailed simulations can be found in Supplemental Material.}
    \label{fig:free_collective}
\end{figure}
\section{Collective dynamics}
For positive alignment strength \(A>0\), the model displays distinct collective dynamics [Fig.~\ref{fig:free_collective}(a)]. Agents self-organize into coherently moving flocks, while transient milling patterns can also arise. After a brief transient, most particles move close to the limiting speed \(v_{\max}\), except during collisions and reorientation events. The boundary potential confines the dynamics to a finite circular domain and smoothly redirects flocks at the edge. A small number of agents may detach from a flock, move independently for some time, and later merge with another group. As a result, the long-time dynamics remain structured and time dependent, with multiple flocks continuously forming, propagating, and reorganizing. Because this interaction produces flock-level motion through an orientation-weighted coupling, we refer to it in what follows as the Flock-Level Orientation-Weighted interaction, or FLOW.
\begin{figure*}
\includegraphics[width=\linewidth]{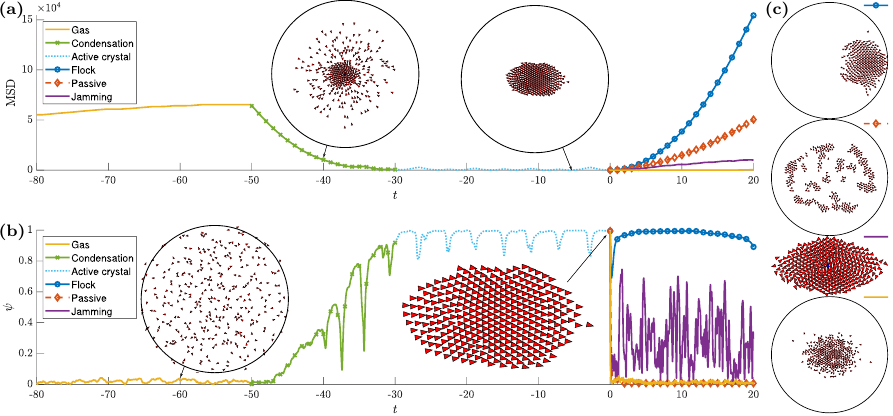}
  \caption{
Representative dynamical protocol for the orientation-weighted active-particle model. The initial condition consists of randomly distributed agents with random velocity directions and speeds. Initially, the system is in the gas-like regime with \(A=-2\). Panels (a) and (b) show the time dependence of the mean squared displacement and polar order parameter, respectively. At \(t=-50\), the anisotropic target field [Eq.~(\ref{eq:a_target})] is switched on, and the alignment strength is changed to \(A=2\). The agents then condense into a dense ordered cluster near the target. At \(t=0\), the parameters \(A\) and \(C\) are changed again, producing the post-target regimes shown in panel (c): flocking, active gas, alignment-free expansion, and jamming. Insets show representative configurations at the times marked by arrows. The distinct behavior of the MSD and \(\psi\) distinguishes the corresponding dynamical regimes. Movies showing the full time evolution together with the MSD and \(\psi\) are provided in the Supplemental Material.
}
    \label{fig:MSD}
\end{figure*}

For \(A<0\), coherent flocks do not persist, and the system evolves toward a gas-like regime [Fig.~\ref{fig:free_collective}(b)]. As discussed above, counter-alignment causes repeated pairwise scattering events, which destroy collective order. As a result, the agents become distributed nearly uniformly throughout the interior region and move predominantly as individuals rather than as coherent groups.

To examine how the orientation-weighted dynamics respond to a localized anisotropic cue, we introduce the target interaction
\begin{equation}
\label{eq:a_target}
\mathbf{a}_i^{\mathrm{tr}} =
C\,\hat{\boldsymbol{\tau}}_i
\left|\hat{\boldsymbol{\tau}}_i \cdot \hat{\boldsymbol n}_t\right|
\Theta(R_t-\tau_i),
\end{equation}
where \(C\) sets the strength of the target interaction, \(\boldsymbol{\tau}_i=\mathbf r_t-\mathbf r_i\), \(\tau_i=|\boldsymbol{\tau}_i|\), and \(\hat{\boldsymbol{\tau}}_i=\boldsymbol{\tau}_i/\tau_i\). Here \(\mathbf r_t\) denotes the target position, \(R_t\) is the interaction range, and \(\hat{\boldsymbol n}_t\) is a unit vector specifying the orientation of the target field.

The target term is constructed as a controlled counterpart of the alignment interaction in Eq.~(\ref{eq:a_fl}). The interparticle direction \(\hat{\mathbf r}_{ij}\) is replaced by the target-centered direction \(\hat{\boldsymbol{\tau}}_i\), while the neighbor velocity \(\mathbf v_j\) is replaced by the prescribed orientation \(\hat{\boldsymbol n}_t\). The target thus acts as a localized oriented source that retains the geometrical projection structure of the interparticle rule.

The absolute value in Eq.~(\ref{eq:a_target}) removes the sign dependence of the projection. In Eq.~(\ref{eq:a_fl}), this sign determines whether the acceleration points toward or away from a neighboring particle along the interparticle axis. Here, instead, the acceleration always points toward \(\mathbf r_t\) within the range \(R_t\). The anisotropy is retained: the inward acceleration is strongest along the target orientation \(\hat{\boldsymbol n}_t\) and vanishes in the perpendicular direction. This construction provides a minimal way to examine how a localized anisotropic source draws agents toward the target and reorganizes the collective motion generated by the orientation-weighted dynamics.

We used the target term to probe how localized anisotropic forcing reorganizes the collective dynamics generated by the alignment interaction. In the main text, the target is stationary, with fixed position and orientation,
\(\mathbf r_t=(0,0)\) and \(\hat{\boldsymbol n}_t=(0,1)\).
For a moving target, the position becomes time dependent, \(\mathbf r_t=\mathbf r_t(t)\), and the orientation of the target field is taken to be parallel to its instantaneous velocity,
\(\hat{\boldsymbol n}_t=\dot{\mathbf r}_t/|\dot{\mathbf r}_t|\).
The corresponding results are presented in the Supplemental Material.

The target field is introduced here as an external localized forcing used to probe how the intrinsic alignment dynamics reorganize under anisotropic confinement and attraction; the flocking and gas-like regimes discussed above, by contrast, arise already in the absence of this forcing.

The collective dynamics are characterized by the mean squared displacement,
\begin{equation}
    \label{eq:MSD}
    \mathrm{MSD} = \frac{1}{N}\sum_{i=1}^N \left|\mathbf{r}_i(t)-\mathbf{r}_i(0)\right|^2,
\end{equation}
and by the polar order parameter
\begin{equation}
    \label{eq:psi}
    \psi = \frac{1}{N v_{\max}} \left| \sum_{i=1}^N \mathbf{v}_i \right|.
\end{equation}
The parameter \(\psi\) measures the degree of collective alignment: \(\psi=1\) when all agents move in the same direction, while \(\psi\to 0\) for disordered motion. This definition is particularly convenient here because the speed-saturation rule keeps most agents moving close to \(v_{\max}\).

To quantify how different parameters affect the ordered state, we consider the local sixfold orientational order parameter
\begin{equation}
    \label{eq:psi6}
    \psi_6(\mathbf{r}_k)
    =
    \frac{1}{N_k}\sum_{l=1}^{N_k}\exp(6 i \theta_{kl}),
\end{equation}
where \(N_k\) is the number of Voronoi nearest neighbors of the \(k\)-th agent, and \(\theta_{kl}\) is the angle between \(\mathbf{r}_{kl}\) and the horizontal axis. The corresponding orientational correlation function is
\begin{equation}
    \label{eq:g6}
    g_6(r)
    =
    \left\langle
    \psi_6^*(\mathbf{r}_i)\psi_6(\mathbf{r}_j)
    \right\rangle \Big|_{r_{ij}\approx r}.
\end{equation}
In the numerical evaluation, the continuous distance \(r\) is discretized into bins of width \(\Delta r=20\), and the average is taken over all pairs whose separations fall into the corresponding bin.

We performed exploratory simulations over a broad range of system sizes and model parameters in order to identify robust dynamical regimes and select representative protocols for detailed analysis. The simulations included two-particle tests (\(N=2\)) and many-body systems with up to \(N=2000\) agents. We varied the alignment strength \(A\), the target strength \(C\), the Lennard--Jones parameters \(r_0\) and \(U_p\), and the integration time step \(\Delta t\). These tests showed that the qualitative organization of the dynamics is controlled primarily by two parameters: the sign of \(A\), which separates alignment-supporting motion from dispersive scattering, and the target strength \(C\), which controls the degree of localization and compression near the anisotropic source. The results presented below use a representative parameter set that displays the main regimes in a single dynamical protocol.
Figure~\ref{fig:MSD}  shows the corresponding protocol and the main sequence of collective regimes discussed below.
For this example, we use the parameters \(N=331,\; r_{\mathrm{cut}}=60,\; r_0=20,\; U_p=33.33,\; U_{\mathrm{shift}}=0.0914, \; k=1,\; R_b=500,\; |A|=2,\; v_{\mathrm{max}}=20,\; \Delta t=0.01,\; R_t=500\).
 Initially, the system is in the gas-like regime with \(A=-2\). When the target field is introduced at the center of the domain ($C=800$) and the regime is switched ($A=2$), agents begin to move inward and accumulate near the target. The resulting dense cluster then starts to oscillate while gradually incorporating the remaining agents. Once the whole system condenses into a single ordered cluster, the dynamics become nearly periodic, with motion along the axis perpendicular to the target orientation. During this stage, the order parameter remains close to unity, with sharp dips marking the short intervals in which the cluster reverses its direction of motion. 
 The movies in the Supplemental Material show that the dense ordered state undergoes coherent back-and-forth reversals over many oscillation periods, with only short reorientation intervals rather than repeated loss and recovery of collective order.

This dense ordered state is related to active crystalline and solid-like phases discussed in earlier theoretical and computational studies. Ordered moving structures of self-propelled particles have been reported in cohesive flocking models with short-range attraction and repulsion, in elastic-sheet descriptions of active solids, and in continuum theories of traveling active crystals~\cite{Gregoire2003,Ferrante2013ActiveCrystals,MenzelLoewen2013,Menzel2014}. In the present model, a comparable type of dense ordered motion arises from a different microscopic mechanism, namely, an orientation-weighted alignment rule combined with short-range interactions and localized forcing.

The back-and-forth motion of the ordered cluster under the anisotropic target field is reminiscent of collective reversals observed in biological groups. In confined fish schools and marching insect bands, local changes in heading or speed can trigger abrupt U-turns of the entire group, even though all individuals obey the same interaction rules and no explicit hierarchy is present \cite{Zheng2024BOC,Escudero2010DirectionalSwitching}. A similar mechanism appears in our model. Because peripheral agents along the target orientation experience strong target-induced acceleration, but weak FLOW interaction they can reverse the motion of the whole cluster while preserving its internal order. In this sense, the observed reversal dynamics provide a minimal active-matter analogue of collective direction changes in animal groups and in swarms responding to localized environmental cues \cite{vanderVaart2019SwarmSpectroscopy}. 
What matters here is not only the presence of a localized cue, but also the geometry of the alignment rule, which makes the outskirt part of the group along target's orientation disproportionately important for the onset of collective turning.

At this point, the target field is removed, which we define as \(t=0\), and the subsequent evolution is examined for several parameter choices shown in Fig.~\ref{fig:MSD}(c).

\emph{Flocking.}  
For positive alignment strength ($A=2$), the agents reorganize into a freely moving flock that preserves the direction acquired before the target is switched off. This regime is characterized by a large and nearly constant order parameter, together with a rapid growth of the mean squared displacement.

\emph{Alignment-free regime.}  
When the alignment interaction is switched off (\(A=0\)), the order parameter remains close to zero and varies only weakly in time. The mean squared displacement increases even faster than in the flocking case. This behavior reflects the strongly compressed initial state at \(t<0\): once the target field is removed, the short-range Lennard-Jones repulsion rapidly expels particles from the dense cluster and redistributes them into passive aggregates. These aggregates then spread approximately isotropically toward the boundary, while some particles remain near the center. In this sense, the alignment-free case provides a passive baseline for comparison with the orientation-weighted dynamics.

\emph{Jamming.}  
If the target field is switched on again with stronger attraction (\(C=4000\), $A=2$) the system enters a jammed regime. In this case, attraction toward the target dominates over alignment, so particles accumulate rapidly in a confined region without developing coherent collective motion. The resulting state is characterized by high local density, strong interparticle repulsion, irregular motion with frequent collisions, suppressed mean squared displacement, and pronounced fluctuations of the order parameter.

\emph{Active gas.}  
For negative alignment strength ($A=-2$), particles disperse throughout the domain, driving the system into a gas-like state. Accordingly, the mean squared displacement exhibits a distinct transition from early-time diffusion to late-time ballistic motion. This crossover reflects a fundamental shift in dynamics: initial frequent interactions randomize trajectories, while increasingly isolated particles later undergo unhindered, ballistic self-propulsion toward the boundaries.

The results above show the dynamical consequences of using a force-based, geometry-dependent alignment rule rather than a fixed-speed angular update. In standard Vicsek-type formulations, alignment directly changes particle headings, while the speed is usually fixed by construction. Here, by contrast, alignment enters as an acceleration term, so steering, speed variation, interparticle collisions, and spatial organization are coupled within the same equations of motion. This velocity and geometry-dependent coupling allows the same microscopic rule to support dispersive dynamics, coherent flocking, dense ordered motion, and collective reversals under localized anisotropic forcing.

We identify these dynamical states using complementary diagnostics rather than by assigning sharp thermodynamic phase boundaries. The mean squared displacement and polar order parameter distinguish flocking, gas-like, jammed, and alignment-free behavior, while the dense ordered state is further characterized by the orientational correlation function \(g_6(r)\), shown in Fig. \ref{fig:orientational_order}. Within the parameter range explored here, the organization of the observed dynamics is governed by simple control mechanisms: the sign of \(A\) determines whether the alignment rule supports coherent motion or dispersive scattering, while the target strength \(C\) controls the degree of localization and compression induced by the anisotropic cue. Increasing \(C\) therefore drives a crossover from coherent oscillatory motion to increasingly confined, disordered, jammed dynamics. For this reason, we focus on diagnostic signatures of representative regimes and their crossovers, rather than on constructing a sharp phase-boundary diagram.
\subsection*{Robustness to angular noise, stochastic kicks, and effective friction}

The collective regimes identified above were obtained in an idealized deterministic setting. Since real active systems are always subject to fluctuations, dissipation, or imperfect response, it is important to test the robustness of these regimes against perturbations of different types. To this end, we consider two complementary modifications of the dynamics. We begin with a standard angular noise of Vicsek type, which perturbs only the direction of motion and therefore provides a natural benchmark for comparison with established active-matter models~\cite{Vicsek1995}. We then introduce a second perturbation that combines stochastic kicks with effective friction and can modify both the direction and the magnitude of the velocity.

As a first test, after the velocity update in Eq.~(\ref{eq:Verlet3}), the velocity direction is additionally rotated by a random angle \(\Delta\theta\in[-\eta/2,\eta/2]\) after speed rescaling, as illustrated in Fig.~\ref{fig:vmax}(b). This perturbation $\hat{R}_{\Delta \theta}$ leaves the speed unchanged and affects only the direction of motion, in direct analogy with the standard Vicsek-type noise~\cite{Vicsek1995}. We study this noise in the dense ordered regime in the presence of the target field (see Supplemental Material). For weak noise, the ordered collective motion remains dominant, and the oscillatory dynamics persist with only modest fluctuations of the order parameter. For strong noise, \(\eta \gtrsim \pi/2\), coherent collective motion is largely destroyed: the agents form a fluctuating gas-like cloud that oscillates as a whole near the target, the average order parameter is reduced, and its fluctuations increase.

As a second test, we supplement Eq.~(\ref{eq:a_tot}) with the term
\begin{equation}
\label{eq:a_i_modified}
    \mathbf{a}_i^{B}
    =
    -B\left[p_{\rm fr}\mathbf{v}_i + p_{\rm t}\boldsymbol{\delta}_i\right],
\end{equation}
where \(B\) sets the overall strength of the perturbation, \(p_{\rm fr}\) and \(p_{\rm t}\) control the relative weights of the frictional and stochastic contributions, and \(\boldsymbol{\delta}_i\) is a random vector drawn from a normal distribution \(N(0,v_{\max}^2)\). In the simulations we use $C=200,$ $B=4.1,$ \(p_{\rm fr}=0.1\) and \(p_{\rm t}=0.8\). At each time step, an agent is assigned either to the alignment dynamics or to the stochastic-friction dynamics. In the latter case, the alignment contribution is temporarily suppressed and \(\mathbf a_i^B\) is applied; otherwise \(\mathbf a_i^B=0\) and the usual alignment term is retained.

Unlike the angular noise, this perturbation affects not only the direction of motion but also the speed. It is intended to model the motion of agents when coherent alignment is weak or absent. In particular, isolated particles no longer move ballistically over long distances but instead undergo noisy, weakly dissipative motion. This provides a simple way to test whether the collective regimes found above survive when deterministic free-flight motion is replaced by a more realistic stochastic background.

\begin{figure}[htb]
\includegraphics[width=\linewidth]{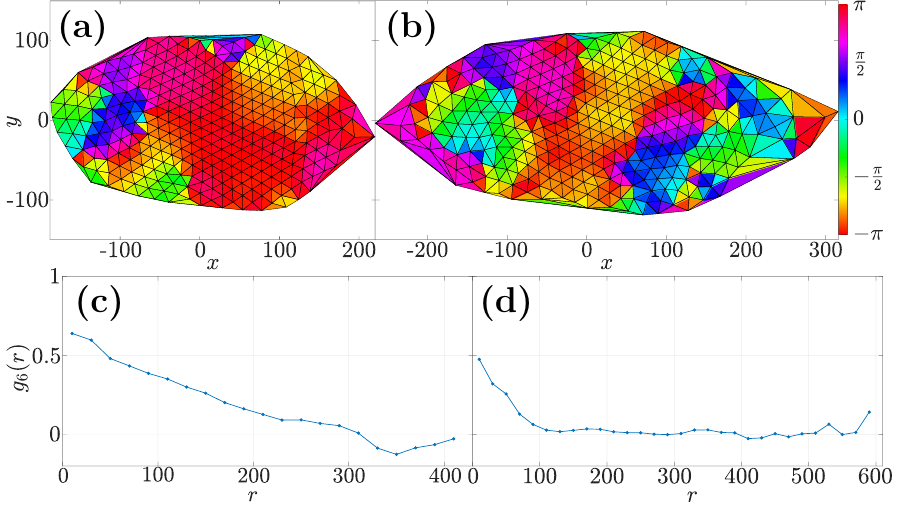}
    \caption{
Local sixfold order and orientational correlation function for the dense ordered state. Panels (a,c) correspond to the unperturbed state shown in Fig.~\ref{fig:MSD} at \(t=0\), while panels (b,d) show the state obtained with the stochastic-friction perturbation of Eq.~(\ref{eq:a_i_modified}). Lines connect Voronoi nearest neighbors. Triangles are colored according to the bond-orientational phase
\(
\varphi=\arg\frac{1}{3}\sum_{i}\psi_6(\mathbf r_i),
\)
where the sum is taken over the three vertices of each triangle. The colorbar indicates the values of \(\varphi\).
}
\label{fig:orientational_order}
\end{figure}

Direct simulations show that the main collective effects remain qualitatively robust under this modification (see Supplemental Material). Flocking and dense ordered motion persist, while the most visible change occurs in the motion of isolated agents, which now wander stochastically and occasionally slow down instead of propagating along nearly straight ballistic trajectories. The effect of this perturbation on the dense ordered state is shown in Fig.~\ref{fig:orientational_order}. Using the orientational correlation function \(g_6(r)\), we find that the ordered structure survives but its orientational correlations decay more rapidly than in the unperturbed case. Thus, stochastic kicks and dissipative slowing weaken the ordered state without immediately destroying it.

Thus, these results show that weak and moderate perturbations do not destroy the main collective regimes, although they progressively reduce orientational order and enhance fluctuations. Strong angular noise eventually suppresses coherent reversals and replaces the dense ordered state by a weakly organized oscillating cloud near the target.

\vspace{1 cm}
\section{Conclusions}

We have studied an agent-based model of self-propelled particles in which alignment is implemented as a velocity-dependent acceleration along the line connecting neighboring particles. This orientation-weighted rule couples the direction and speed of a neighbor to its relative position, thereby linking alignment, spatial organization, and acceleration within a single microscopic interaction. In the absence of external forcing, changing the sign of the alignment strength separates two qualitatively different behaviors: positive alignment supports coherent flocking, whereas negative alignment leads to dispersive gas-like motion.

We then examined how the same dynamics responds to a localized anisotropic cue constructed as a controlled counterpart of the orientation-weighted interaction. This target field is a special stationary agent with stronger and symmetrical attraction, whereas the geometrical weighting of the interparticle interaction is retained. This forcing gathers agents and reorganizes their collective motion. Depending on the target strength and alignment parameters, the system develops either a densely ordered moving cluster with crystal-like local order and coherent, nearly periodic reversals, or a jammed high-density state with strong fluctuations and suppressed transport.

The collective regimes were characterized using complementary dynamical and structural diagnostics, including the mean squared displacement, the polar order parameter, the local sixfold order parameter, and the orientational correlation function. These diagnostics show that the dense ordered state retains persistent bond-orientational order, while stronger target forcing or increasing noise progressively weakens coherent motion and orientational correlations. The main behaviors also persist under angular noise and under stochastic perturbations with effective friction, although fluctuations reduce order and eventually suppress coherent reversals.

These results show that orientation-weighted, geometry-dependent alignment provides a flexible microscopic mechanism for reorganizing collective motion. By linking the local response of each agent to relative position, neighbor velocity, and finite propulsion capacity, the model captures how simple interaction rules can generate both spontaneous collective motion and structured responses to localized anisotropic forcing.

\section*{Acknowledgments}
We thank A. O. Oliinyk and Y. O. Nikolaieva for
useful discussions.

\bibliography{Refs}

\end{document}